\documentclass[%
 reprint,
superscriptaddress,
 amsmath,amssymb,
 aps,
]{revtex4-2}

\usepackage{graphicx}
\usepackage{dcolumn}
\usepackage{bm}

\begin{document}

\preprint{APS/123-QED}

\title{Polymer identification via undetected photons using a low footprint nonlinear interferometer}
\author{Atta Ur Rehman Sherwani}
\affiliation{Institut f{\"u}r Physik, Humboldt-Universit{\"a}t zu Berlin, Newtonstr. 15, 12489 Berlin, Germany}

\author{Emma Pearce}
\email{emma.pearce@glasgow.ac.uk}
\affiliation{Institut f{\"u}r Physik, Humboldt-Universit{\"a}t zu Berlin, Newtonstr. 15, 12489 Berlin, Germany}
\affiliation{School of Physics and Astronomy, University of Glasgow, Glasgow, G12 8QQ, UK}

\author{Philipp Hildenstein}
\affiliation{Ferdinand-Braun-Institut (FBH), Gustav-Kirchhoff-Straße 4, 12489 Berlin, Germany}

\author{Felix Mauerhoff}
\affiliation{Ferdinand-Braun-Institut (FBH), Gustav-Kirchhoff-Straße 4, 12489 Berlin, Germany}

\author{Alexander Sahm}
\affiliation{Ferdinand-Braun-Institut (FBH), Gustav-Kirchhoff-Straße 4, 12489 Berlin, Germany}

\author{Katrin Paschke}
\affiliation{Ferdinand-Braun-Institut (FBH), Gustav-Kirchhoff-Straße 4, 12489 Berlin, Germany}

\author{Helen M. Chrzanowski}
\affiliation{Institut f{\"u}r Physik, Humboldt-Universit{\"a}t zu Berlin, Newtonstr. 15, 12489 Berlin, Germany}

\author{Sven Ramelow}
\email{sven.ramelow@physik.hu-berlin.de}
\affiliation{Institut f{\"u}r Physik, Humboldt-Universit{\"a}t zu Berlin, Newtonstr. 15, 12489 Berlin, Germany}
\affiliation{Ferdinand-Braun-Institut (FBH), Gustav-Kirchhoff-Straße 4, 12489 Berlin, Germany}

\date{\today}

\begin{abstract}

Plastic pollution has become a critical global challenge, with microplastics pervading ecosystems and entering human food chains. Effectively monitoring this widespread contamination demands rapid, reliable, and portable material identification techniques that often elude conventional Raman and FTIR spectroscopy. Undetected photon spectroscopy within a nonlinear interferometer (NLI) offers a solution, allowing the retrieval of mid-infrared absorption spectra by detecting only near-infrared signal photons using standard silicon-based technology. 
Here, we demonstrate a highly compact, micro-integrated, thermally-stabilised NLI with a Michelson-like geometry designed for the rapid spectroscopy of plastics. We benchmarked its room-temperature performance, demonstrating a signal-to-noise ratio of 34 with a measurement rate of 100 Hz and a spectral resolution of 6 cm$^{-1}$. We show that we can accurately and rapidly retrieve the characteristic vibrational absorption spectra of common polymers such as polypropylene, polyethene, and polystyrene, without using mid-infrared technology. These results establish our compact module as a promising field-deployable platform for robust, real-time environmental monitoring systems and other mid-infrared spectroscopy applications.
\end{abstract}

\maketitle


\section{\label{sec:introlevel1}Introduction}

Plastic pollution, encompassing both macro- and microplastics, is a growing environmental and health crisis\cite{leslieDiscoveryQuantificationPlastic2022,zhaoMicroplasticHumanDietary2024,liPotentialHealthImpact2023}. Microplastics originate from diverse sources, including textile fibre abrasion, tyre wear, and the fragmentation of larger debris. These particles permeate marine, terrestrial, and atmospheric systems, accumulating in food chains and human tissues \cite{thompsonTwentyYearsMicroplastic2024, wrightPhysicalImpactsMicroplastics2013, nihartBioaccumulationMicroplasticsDecedent2025}. Despite the urgency of this issue, current monitoring methods remain too slow, expensive, and laboratory-bound to support continuous, large-scale surveillance.

Established analytical techniques, notably Raman spectroscopy, Fourier-transform infrared spectroscopy (FTIR), and pyrolysis gas chromatography mass spectrometry (Py-GC/MS), are widely used for polymer identification but face significant practical limitations in the field\cite{picoPyrolysisGasChromatographymass2020,araujoIdentificationMicroplasticsUsing2018,kapplerAnalysisEnvironmentalMicroplastics2016,schymanskiAnalysisMicroplasticsWater2018}. Raman spectroscopy offers chemical specificity but often suffers from fluorescence interference and low signal-to-noise ratios (SNR) in complex environmental samples. While FTIR enables chemical fingerprinting, it is diffraction-limited to particles larger than approximately 20~\textmu m and struggles with water absorption in environmental matrices. Finally, Py-GC/MS is destructive and time-intensive, rendering it unsuitable for in-situ monitoring.

Nonlinear interferometry (NLI) with undetected photons has recently emerged as a powerful alternative for infrared spectroscopy \cite{kaufmannMidIRSpectroscopyNIR2022, paterovaNonlinearInfraredSpectroscopy2017,lindnerFourierTransformInfrared2020,lindnerHighsensitivityQuantumSensing2023,kuritaQuantumInfraredAttenuated2025,tashimaUltrabroadbandQuantumInfrared2024,hashimotoBroadbandSpectroscopyInterferometry2025,hashimotoFouriertransformInfraredSpectroscopy2024,wornerUltrafastSpectroscopyLiquids2025,nevesOpenpathDetectionOrganic2024,kalashnikovInfraredSpectroscopyVisible2016,lindnerNonlinearInterferometerFouriertransform2021} and sensing \cite{lemosQuantumImagingUndetected2014,defienneAdvancesQuantumImaging2024,kviatkovskyMidinfraredMicroscopyPosition2022,kviatkovskyMicroscopyUndetectedPhotons2020,gilabertebassetPerspectivesApplicationsQuantum2019,leon-torresMidInfraredQuantumScanning,kutasTerahertzQuantumSensing2020,haasePhasequadratureQuantumImaging2023,pearceSingleframeTransmissionPhase2024,pearcePracticalQuantumImaging2023,topferQuantumHolographyUndetected2022,dongMethaneSensingUnbalanced2025,gemmellCouplingUndetectedSensing2024,plackeMidIRHyperspectralImaging2026}. In this technique, correlated near-infrared (NIR) signal and mid-infrared (MIR) idler photon pairs are generated via spontaneous parametric down-conversion (SPDC). The idler photons probe the molecular vibrations of the sample, while only the signal photons are detected. Through quantum interference, the absorption experienced by the undetected idler is mapped on to the near-infrared signal. This allows access to the mid-infrared absorption spectrum using mature, high-performance silicon-based detection technology. The result is a system that offers higher SNR, faster acquisition, and a smaller footprint compared to traditional mid-infrared systems that rely on noisy or cryogenically-cooled detectors.

Building upon these developments, we present a modular NLI designed specifically for robust, rapid plastic identification in the 3060–2790 cm$^{-1}$ region. Unlike table-top experimental setups, our interferometer is fully integrated into a micro-optical housing that ensures a small form factor along with mechanical and thermal stability. We characterise the system's noise performance, demonstrating a high SNR (up to 34 operating at 100 Hz) and a spectral resolution of 6 cm$^{-1}$ sufficient to resolve the C--H absorption bands of common polymers. To validate its utility, we retrieve the mid-infrared absorption spectra of polypropylene (PP), polyethylene (PE), and polystyrene (PS) films. These results highlight the potential of our modular NLI as a compact, alignment-stable platform for next-generation field-deployable environmental monitoring sensors.

\section{Experiment}

\subsection{Nonlinear interferometer}

\begin{figure}[htbp]
\centering
\includegraphics[width=\columnwidth]{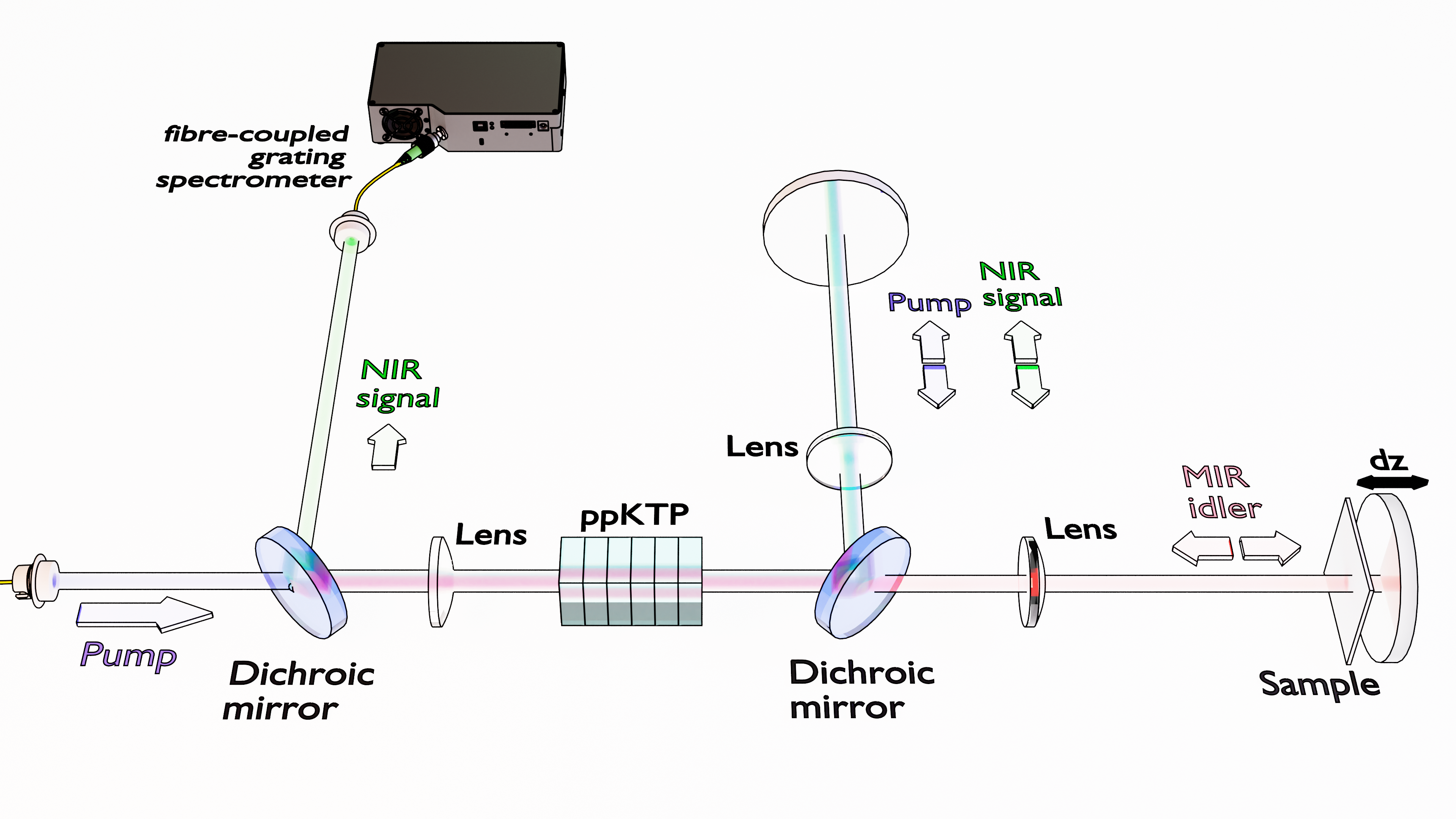}
\caption{Schematic of the Michelson-type NLI. A pump laser (720\,nm, shown in blue) is focused into a nonlinear crystal (ppKTP), generating correlated NIR signal (901\,-\,923\,nm, shown in green) and MIR idler photons (3.26\,-\,3.58\,\textmu m, shown in red) via SPDC. The pump and signal are separated from the idler by a dichroic mirror. The idler interacts with the sample in one path while the pump and signal traverse a separate path, before all fields are reflected back through the crystal. Absorption or phase shifts experienced by the idler field are mapped onto an interference pattern detected in the signal, which is collected by a fibre-coupled grating spectrometer.}
\label{fig:NLI_schematic}
\end{figure}

Our NLI module operates on the principle of spontaneous parametric down-conversion (SPDC) and induced coherence without induced emission\cite{zouInducedCoherenceIndistinguishability1991,chekhovaNonlinearInterferometersQuantum2016}, implemented in a compact, Michelson-type geometry (Figure~\ref{fig:NLI_schematic}). A 720\,nm pump laser is focused into a periodically-poled potassium titanyl phosphate (ppKTP), generating spectrally and temporally correlated photon pairs, a NIR signal (901\,--\,923\,nm) and MIR idler photons (3.26\,--\,3.58\,\textmu m). This MIR range is specifically engineered to cover the spectral absorption region of many polymers. To ensure broadband operation, the group velocity dispersion conditions in the nonlinear crystal are matched for the signal and pump velocities, while periodic poling provides the necessary quasi-phase-matching (QPM) to satisfy momentum conservation \cite{vanselowUltrabroadbandSPDCSpectrally2019}. Following generation, a dichroic mirror splits the optical path; the MIR idler is transmitted towards the sample, while the pump and NIR signal are reflected into a separate path. Mirrors at the end of both paths reflect the fields back through the crystal for a second pass, a configuration that minimizes alignment complexity and provides excellent passive stability by sharing common optics.
The sensing mechanism relies on the interaction of the idler with the sample. As the idler traverses the sample, it experiences absorption and phase shifts, which are mapped onto the signal field during the second pass. The resulting signal interference spectrum is then retrieved using a fibre-coupled grating spectrometer. Fundamentally, quantum interference occurs only when the photon states from the first and second passes are indistinguishable. If any "which-path" information is introduced for example, by absorption in the idler path the indistinguishability is broken, and quantum interference is inhibited \cite{lemosQuantumImagingUndetected2014,zouInducedCoherenceIndistinguishability1991}.

The interference is characterised by its visibility, defined as:

\begin{equation}
    \mathrm{VIS} = \frac{(I_{\max} - I_{\min})}{(I_{\max} + I_{\min})} \approx \frac{2 \sqrt{I_1 I_2}}{I_1 + I_2} |\tau_i|^2,
\label{eq:vis_equation}
\end{equation}
where $I_{\max}$ and $I_{\min}$ denote the maximum and minimum intensities of the interference fringes, and $I_1$ and $I_2$ represent the independent spectral intensities of the signal photons generated during the first and second passes, respectively. The term $|\tau_i|^2$ represents the round-trip absolute amplitude transmission of the idler field passing through the sample twice, which is directly equivalent to the sample's standard intensity transmittance ($T$). High visibility indicates excellent spatial and spectral overlap between the two generation processes, whereas reduced visibility reflects decoherence, misalignment, unequal path balancing, or crucially, absorption in the sample. Absorption does not reduce the generation rate of the second pass ($I_2$); rather, it attenuates the returning idler field. This loss of idler photons provides "which-path" information to the environment, destroying the quantum indistinguishability between the two generation events and directly reducing the interference visibility \cite{zouInducedCoherenceIndistinguishability1991}.

\subsection{Single-shot spectrum retrieval (envelope approach)}

\begin{figure*}[htbp]
\centering
\includegraphics[width=0.8\linewidth]{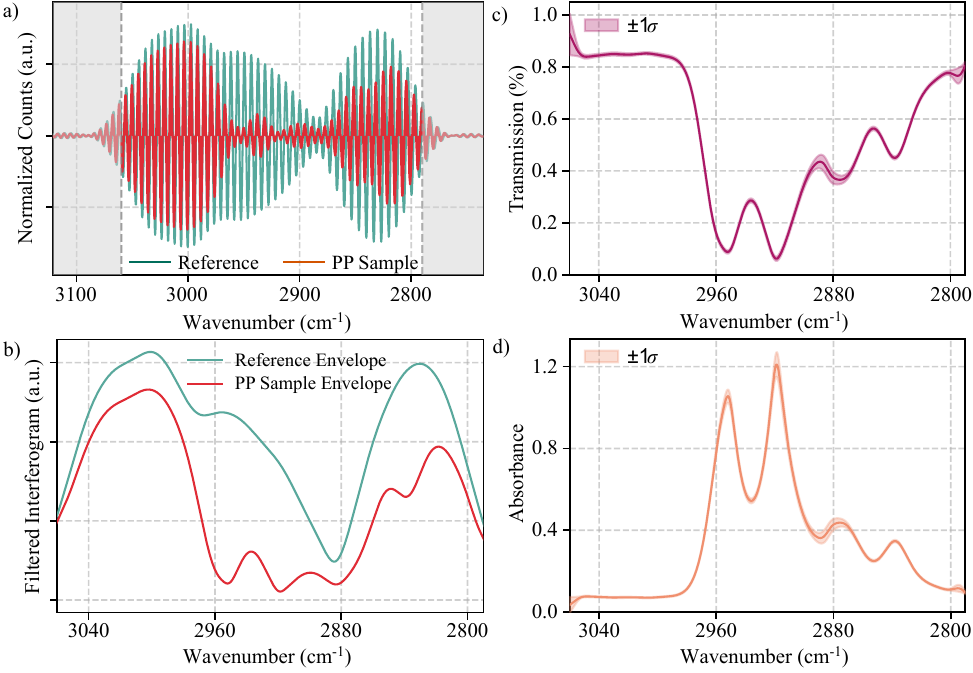}
\caption{Measurement and data-extraction procedure for a polypropylene (PP) film using our single-shot scheme (a) A reference interferogram taken without the sample (dark cyan) and an interferogram taken with the sample (red). The shaded region indicates the region omitted for further processing (b) Envelopes extracted via a Hilbert transform. (c) Transmission spectrum from the ratio of the sample to the reference envelope; shaded region indicates $\pm1\sigma_{noise}$. (d) Absorbance spectra calculated via the Beer-Lambert law.}
\label{fig:Absorption_extraction_sequence}
\end{figure*}

We employ an envelope-based method to retrieve the absorbance spectrum from a single interferogram \cite{kaufmannMidIRSpectroscopyNIR2022}. In this scheme, a carrier frequency is introduced by creating an optical path length difference (OPLD) between the signal and idler paths, where the carrier modulation depth (visibility) is directly proportional to the sample's transmission. By setting a sufficiently large OPLD, the spectral information encoded in the carrier frequency can be separated from a low-frequency background, provided that the carrier frequency can be resolved by the spectrometer\cite{shannonCommunicationPresenceNoise1949,leitgebPerformanceFourierDomain2003}.

Figure~\ref{fig:Absorption_extraction_sequence} illustrates the measurement and data extraction sequence for a polystyrene (PS) film. First, a reference interferogram (blue) is recorded without the sample, followed by a sample interferogram (red). The interferograms are then resampled to the wavenumber domain and truncated to the usable spectral bandwidth, excluding the noise-dominated regions shown in grey in Figure~\ref{fig:Absorption_extraction_sequence}a. The envelope for each interferogram is extracted using a Hilbert transform implemented via Fourier-domain filtering, which separates the interference term from the DC background (Figure~\ref{fig:Absorption_extraction_sequence}b). The transmission spectrum is then calculated from the ratio of the sample envelope to the reference envelope, effectively isolating the visibility reduction caused by the sample (Figure~\ref{fig:Absorption_extraction_sequence}c). Finally, the transmission is converted to absorbance (Figure~\ref{fig:Absorption_extraction_sequence}d) using the Beer–Lambert relation.

The detected signal intensity $I(\nu_s)$ in the interferometer can be expressed as

\begin{equation}
I(\nu_s) = S(\nu_s) \left[1 + V(\nu_s) \cos\left(2\pi \nu_s \Delta L + 2\delta \phi_i\right)\right],
\label{eq:intensity}
\end{equation}
where $S(\nu_s)$ is the spectral intensity of the signal, $V(\nu_s)$ is the frequency-dependent interference visibility, $\delta \phi_i$ is the single-pass phase shift induced by the sample on the idler, and $\Delta L$ represents the OPLD between the interferometer paths. The cosine term describes the carrier frequency of the interference fringes set by the OPLD in the wavenumber domain. The modulation envelope is extracted by applying a Hilbert transform, $\mathcal{H}\{I(\nu_s)\}$, and taking the magnitude of the resultant analytical signal:

\begin{equation}
E(\nu_s) = |I(\nu_s) + i \mathcal{H}[I(\nu_s)]| \approx S(\nu_s) V(\nu_s).
\label{eq:reconstructed}
\end{equation}
Here, $E(\nu_s)$ represents the extracted envelope, which depends on both the signal spectral intensity and the visibility. The transmission spectrum $T(\nu_s)$, is defined as the ratio between the envelopes recorded with and without the sample. Assuming the signal spectrum remains constant between measurements, it cancels out in the ratio:

\begin{equation}
T(\nu_s) = \frac{E_{\text{sample}}(\nu_s)}{E_{\text{reference}}(\nu_s)} = \frac{S(\nu_s) V_{\text{sample}}(\nu_s)}{S(\nu_s) V_{\text{reference}}(\nu_s)} = \frac{V_{\text{sample}}(\nu_s)}{V_{\text{reference}}(\nu_s)}.
\label{eq:visibility_to_transmission}
\end{equation}
Since the visibility in this double-pass geometry is linearly proportional to the sample's intensity transmission, this ratio yields the absolute transmission $T(\nu_s)$ directly\cite{kaufmannMidIRSpectroscopyNIR2022}. Once the transmission spectrum is obtained, the corresponding absorbance $A(\nu_s)$ is derived using the Beer–Lambert law:

\begin{equation}
A(\nu_s) = -\log_{10}\left[T(\nu_s)\right].
\label{eq:absorbance} 
\end{equation}
Provided the initial interference visibility is sufficient to distinguish the signal from the noise floor, this retrieved absorbance is directly comparable to conventional FTIR or ATR spectra.

\subsection{Module Design and Operation of the Nonlinear Interferometer}

\begin{figure}[htbp]
\centering
\includegraphics[width=\columnwidth]{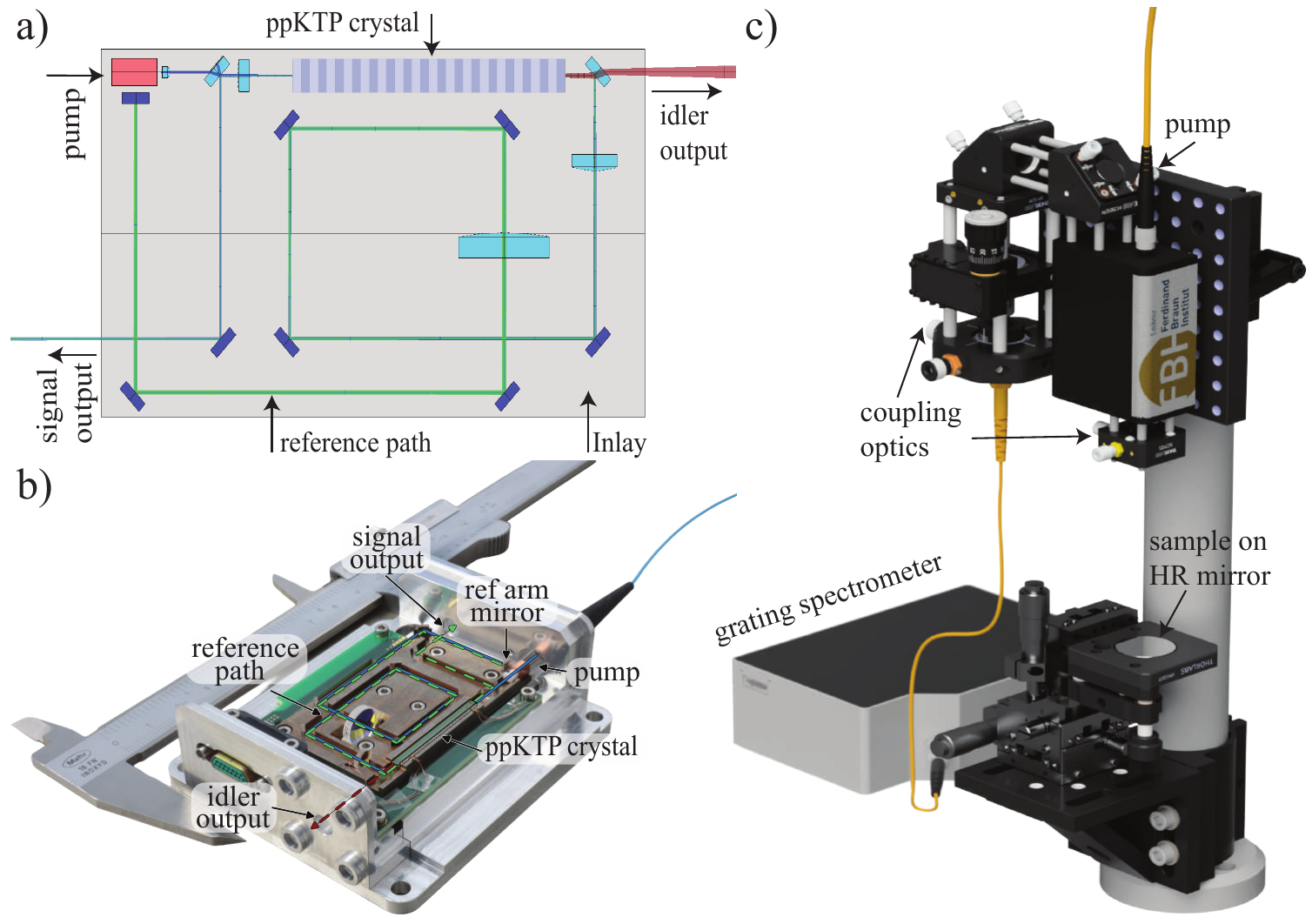}
\caption{Design and experimental configuration of the compact nonlinear interferometer module. (a) Schematic of the integrated micro-optical module showing the internal beam routing. (b) Photograph of the manufactured module housing without the cover. (c) Complete experimental setup including the fibre-coupled module, external signal coupling optics, sample holder, and grating spectrometer.}
\label{fig:module}
\end{figure}

To realise a robust sensing platform, the nonlinear interferometer was integrated into a micro-optical module by Ferdinand-Braun-Institut (FBH) featuring a compact form factor of $95 \times 75 \times 30 $mm$^3$. Figure~\ref{fig:module}a illustrates the internal beam routing. The pump field is injected via a fibre towards the nonlinear crystal, after which a dichroic mirror separates the idler from the signal and pump fields. The idler exits the module to interact with the sample, while the pump and signal are circulated internally within the housing.  This internal reference path extends for approximately 200\,mm, which sets the required working distance for the external idler path to achieve interference. All three fields are reflected and focused back into the crystal for the second pass. Following this second interaction, the signal light is separated from the pump via a second dichroic mirror and coupled into an output fibre for detection. Mechanically, the system features a single monolithic baseplate and integrated thermoelectric cooling (TEC) for temperature stabilization, ensuring that phase-matching conditions are preserved across varying environmental temperatures (Figure~\ref{fig:module}b). 

In the experimental configuration, shown in Figure~\ref{fig:module}c, a continuous-wave Ti:sapphire laser at 720 nm was fibre-coupled into the module and focused into a 30 mm long ppKTP crystal. The crystal was quasi-phase-matched for a collinear type-0 SPDC process to generate highly non-degenerate photon pairs, with the idler covering the 3.26–3.58 \textmu m (3060\,–\,2790\,cm$^{-1}$) spectral range. The idler beam was collimated to a $\sim$\,2.3\,mm diameter using standard 16\,mm cage system optics, which allow for rapid exchange of collimating and focusing elements. The idler was directed onto the sample located at the end mirror of the sample path. The output signal photons (901\,–\,923\,nm) were collected via a single-mode fibre and detected using a silicon-based grating spectrometer (MayaPro, Ocean Optics).

\begin{figure}[htbp]
\centering
\includegraphics[width=\columnwidth]{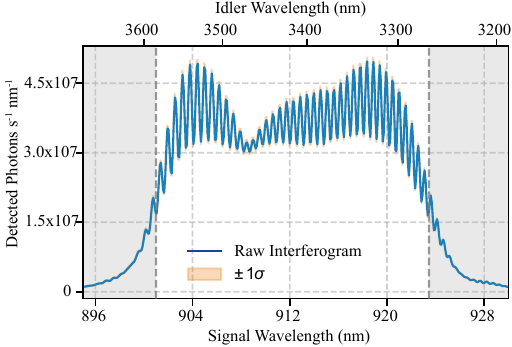}
\caption{Raw SPDC interferogram acquired at a 10\,ms integration time. The blue interferogram shows the acquired spectrum, and the orange shaded band represents the $\pm 1\sigma_\text{noise}$, consistent with the expected Poissonian shot noise of the photon count. The vertical axis shows detected photon flux (converted from spectrometer counts). The 22\,nm (270\,cm$^{-1}$) bandwidth utilised for absorption analysis is indicated by the unshaded central region, while the grey shaded areas fall outside the region of interest.}
\label{fig:rawspectra_photoncounts}
\end{figure}

Figure~\ref{fig:rawspectra_photoncounts} presents a background-subtracted raw interferogram acquired without a sample.  Measurements were performed with 10\,ms integration time and 400\,mW of pump power coupled into the input fibre. Accounting for the effective SPDC bandwidth of 22\,nm in the signal domain (corresponding to 270\,cm$^{-1}$ in wavenumber space), the detector's quantum efficiency ($\sim$42\%), the spectrometer gain (0.35), grating efficiency ($\sim$50\%), and the spectrometer pixel resolution of 0.089\, nm/pixel, we estimate a total photon flux of $3.6 \times 10^{7}$ detected photons per second.
A pronounced spectral dip was observed near 908\,nm, corresponding to an idler wavelength of 3.45\,\textmu m, which arises from intrinsic phosphate absorption within the KTP crystal \cite{hanssonTransmissionMeasurementsKTP2000,morrisInfraredStudyOH1992}. Experimentally, the maximum observed visibility was $\sim$18.5\% under standard operating conditions, dropping to $\sim$5\% at the absorption wavelength. While shorter crystals utilised in previous mid-IR studies exhibited negligible loss \cite{vanselowUltrabroadbandSPDCSpectrally2019, kaufmannMidIRSpectroscopyNIR2022}, the intrinsic idler absorption in the 30\,mm crystal noticeably reduces the overall system visibility in this region.

Since the SNR of the measurement is directly proportional to the interference visibility, the inherent crystal absorption across the bandwidth effectively reduces the sensitivity of the system, particularly around 3.45\,\textmu m. Consequently, while weak absorbers can still be detected, the quantitative accuracy of the absorbance measurement may be compromised in this band, potentially resulting in distorted peak shapes. Strategies to mitigate this effect, such as optimal crystal length selection, are discussed in the Appendix.

\section{Results \& Discussion}

\subsection{System Characterization}

To establish the baseline performance of the micro-integrated NLI module, we conducted a two-stage characterisation process. First, we experimentally evaluated the trade-off between spectral resolution and interference visibility to define an optimal OPLD operation point that balances resolving power with signal detection efficiency. Second, we quantified the sensitivity and stability limits of the system. This involved calculating the SNR from the spectral noise and utilizing Allan-Werle deviation analysis to verify the system's operation in the shot-noise limited regime and to extrapolate potential performance gains.

We first evaluated the trade-off between spectral resolution and interference visibility by varying the OPLD. While increasing the OPLD improves resolution (as explained in Section 2.2), it simultaneously leads to a signal roll-off, which is a characteristic decay in detected contrast. This reduction occurs because the period of the interference fringes decreases as the path length difference grows. Consequently, as the modulation frequency of the fringes approaches the limits of the spectrometer's finite spectral resolution (0.089 nm per pixel), the effective visibility is washed out by spectral averaging \cite{vanselowFrequencydomainOpticalCoherence2020,leitgebPerformanceFourierDomain2003}.

\begin{figure}[htbp]
    \centering
    \includegraphics[width=\columnwidth]{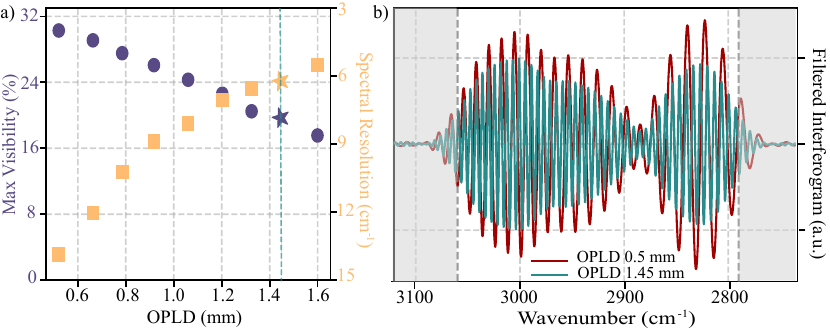}
    \caption{(a) Trade-off between interference visibility (dark purple circles, left y-axis) and spectral resolution (gold squares, right y-axis) as a function of relative optical path length difference (OPLD). The star symbols and the dark cyan vertical dashed line highlight the selected operating point (OPLD\,=\,1.45\,mm), which balances adequate resolution ($\sim$6\,cm$^{-1}$) with visibility. (b)
    Representative filtered interferograms. The measurement at OPLD\,=\,0.5\,mm (red) exhibits high visibility but low spectral resolution, whereas the selected operating point at OPLD\,=\,1.45\,mm (dark cyan) achieves higher resolution at the cost of slightly reduced visibility.}
    \label{fig:Visibility-spectralresolution_plot}
\end{figure}

To analyse this behaviour, the spectral resolution values for each OPLD were derived via simulation. We numerically introduced a narrowband absorption feature (width 0.96 cm$^{-1}$) into the reference data. The transmission spectra were then extracted using the standard envelope method, and the full width at half maximum (FWHM) of the reconstructed feature was determined via Gaussian fitting.
The resulting dependencies are shown in Figure~\ref{fig:Visibility-spectralresolution_plot}(a): increasing the relative OPLD from 0.5 mm to 1.6 mm improves the attainable spectral resolution from 14 cm$^{-1}$ to 6 cm$^{-1}$. Simultaneously, as the finite spectrometer resolution effects begin to dominate, the maximum observed visibility decreases from 30\% to 17.5\%. This effect is illustrated in Figure~\ref{fig:Visibility-spectralresolution_plot}(b) which compares two raw interferograms resulting from different OPLDs: the red interferogram is the result of a short OPLD (0.5 mm), showing slow carrier frequency and high visibility; the dark cyan interferogram with a longer OPLD (1.45 mm) has a high carrier frequency and low contrast. 

Based on this analysis, an operation point was set at an OPLD of 1.45 mm (shown with a star) for subsequent characterisation. At this setpoint, the system achieves a spectral resolution of around 6 cm$^{-1}$, sufficient to resolve the vibrational bands of the target polymers and a peak interference visibility of approximately $\sim$18.5\%. While a longer OPLD could theoretically yield higher resolution, the operating point was practically limited by the intrinsic mid-IR absorption in the KTP crystal. The visibility is not spectrally uniform, showing a distinct minimum of $\sim$5\% at a single strong absorption resonance at 3.45~\textmu m\cite{hanssonTransmissionMeasurementsKTP2000,morrisInfraredStudyOH1992}. Increasing the OPLD further would reduce the visibility in this critical spectral region below 5\%, potentially compromising polymer identification. Thus, 1.45 mm represents the optimal balance for our system, maximising resolution without extinguishing the signal at the crystal's absorption dip. Note from Figure~\ref{fig:Visibility-spectralresolution_plot}(a) that by trading off spectral resolution, one can approximately linearly increase the visibility and thus final measurement speed.

\begin{figure}[htbp]
\centering
\includegraphics[width=\columnwidth]{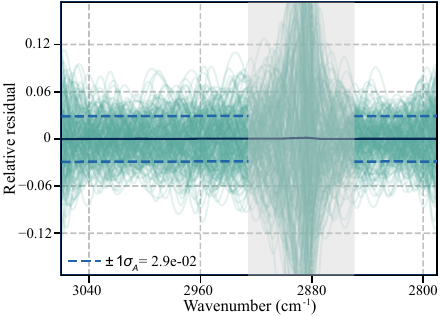}
\caption{Noise analysis of the nonlinear interferometer at 10 ms. The dark cyan trace shows the normalised spectral residuals of 200 spectra calculated from two successive single-channel spectra. The dark blue line in the centre indicates the relative mean residual (near zero), and the blue dashed lines represent the $\pm 1\sigma_A$ bounds where $\sigma_A$ is the Allan deviation. The shaded region between 2850-2925 cm$^{-1}$ indicates interval excluded from analysis due to strong intrinsic crystal absorption.}
\label{fig:Pairwise_spectral_noise}
\end{figure}

\begin{figure}[htbp]
\centering
\includegraphics[width=\columnwidth]{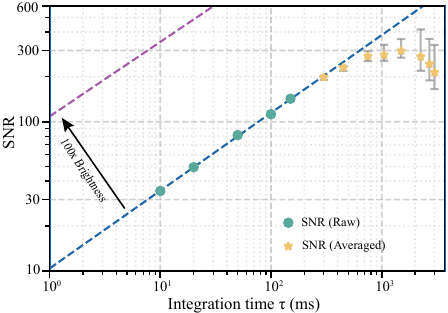}
\caption{Averaged SNR values from spectra acquired with integration times of 10\,-\,150\,ms (dark cyan circles) and for block-averaged spectra up to 3\,s (gold stars). The blue dashed line indicates the spectrometer-detected photon's theoretical shot-noise scaling, following a square-root relation. This ideal scaling can be observed up to $\tau \approx 0.5$ s, after which drift-induced instabilities cause the SNR to saturate and subsequently decline.  The purple dashed line indicates an extrapolated SNR expected from a $\times100$ increase in photon flux, predicting an SNR of $\sim$100 at 1 kHz measurement speeds. }
\label{fig:SNR_scaling_plot}
\end{figure}

To quantify the system's SNR-scaling with integration time, we evaluated its Allan-Werle deviation, $\sigma_{\mathrm{A}}(\tau)$~\cite{ZelingerSignalProcessing2013,werleLimitsSignalAveraging1993,gattingerQuantumFourierTransform2025}. Unlike the standard deviation, which treats all noise mechanisms equally, this metric discriminates between noise types based on their timescale. This allows for the identification of the optimum integration time $\tau$ where the system transitions from a precision-limited white noise regime to a drift-dominated instability regime. At the chosen OPLD, a series of single-spectrum measurements was recorded over a range of integration times. The spectral noise was assessed by calculating the normalised difference between consecutive spectral acquisitions. The scalar Allan-Werle deviation was then computed, for each integration time, over the selected spectral bandwidth according to
\begin{equation}
\sigma_{\mathrm{A}}(\tau) = \sqrt{\frac{1}{2(M-1)} \sum_{k=1}^{M-1} \left( \bar{y}_{k+1}(\tau) - \bar{y}_k(\tau) \right)^2 },
\label{eq:allan_variance}
\end{equation}
where $\bar{y}_{k}(\tau)$ denotes the normalised spectral intensity of the $k$-th acquisition interval averaged over a duration $\tau$, and $M$ is the total number of such intervals in the time series. Figure~\ref{fig:Pairwise_spectral_noise} shows the pairwise spectral residuals obtained for the base integration time of 10~ms with a corresponding Allan-Werle deviation of $\sigma_{\mathrm{A}} \approx 2.9 \times 10^{-2}$, resulting in a single-shot SNR ratio of approximately~34. We define SNR here as $\mathrm{SNR}(\tau) = 1/\sigma_{\mathrm{A}}(\tau)$ and evaluate it for integration times from 10~ms to 150~ms (spectrometer saturation limit) to characterise the scaling of the SNR, as presented in Figure~\ref{fig:SNR_scaling_plot} (denoted by cyan circle). To extend this analysis beyond the saturation limit of the spectrometer, we computed block-averaged values up to 3~s (gold stars in Figure~\ref{fig:SNR_scaling_plot}) by combining consecutive short-integration spectra. 

The result in Figure~\ref{fig:SNR_scaling_plot} shows that in the range of $0.01~\text{s}<\tau<0.5~\text{s}$, the system exhibits a square-root scaling ($\text{SNR} \propto \sqrt{\tau}$). The square-root scaling corresponds to the shot-noise limited regime of the detected photons, without limitations due to detector noise or external instabilities. This is in contrast to the typical linear scaling expected when relative intensity noise is the dominant noise source.Furthermore, the observed scaling confirms that the integration time required for a fixed SNR is inversely proportional to source brightness, motivating straightforward improvement strategies. Future implementations using waveguide enhancement \cite{babelUltrabrightTwocolorPhoton2025} or resonant pump-power enhancement \cite{mannLownoiseQuantumFrequency2023} promise at least $\times$100 brighter sources; their extrapolated performance (purple dashed line in Figure~\ref{fig:SNR_scaling_plot}) suggests a potential SNR exceeding 100 at kHz rates. Beyond integration times of approximately 1.5~s, the SNR saturates and subsequently declines. This turnover indicates the onset of low-frequency drifts on timescales typical for mechanical oscillation, air motion, or thermal effects \cite{werleLimitsSignalAveraging1993}, which begin to dominate over quantum noise at these longer timescales. While shielding the external mid-IR interferometer path may help to mitigate these noise sources, the focus of this work was rapid measurement performance. Given the intended application of field-deployable plastic identification, stability on short timescales is prioritized, making the system's robust millisecond performance highly favourable for live monitoring.

\subsection{Fast spectroscopy of polymers}
\begin{figure*}[htbp]
    \centering
    \includegraphics[width=0.8\linewidth]{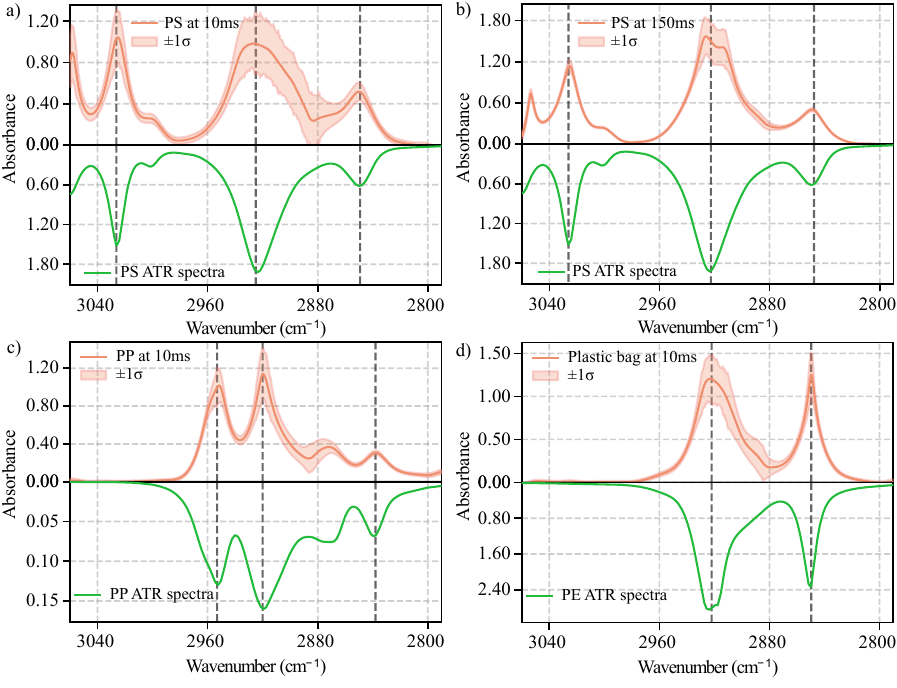}
    \caption{Absorbance spectra of polymer thin films retrieved via the nonlinear interferometer (orange) compared to reference ATR-FTIR spectra (green, inverted). (a) A polystyrene (PS) recorded at 10 ms integration time (100 Hz). (b) The same PS sample recorded at 150 ms integration time (6.6 Hz), showing reduced noise but a peak deformation, which we attribute to signal saturation. (c) A polypropylene (PP) sample with 4~\textmu m thickness. (d) A polyethylene (PE) sample from a commercial shopping bag. The shaded orange regions indicate the $\pm 1\sigma$ noise floor. Vertical dashed lines mark the alignment of characteristic vibrational bands between the NLI measurements and the industrial reference (4 cm$^{-1}$ resolution)}
    \label{fig:Spectra_results}
\end{figure*}

To demonstrate the system's spectroscopic sensing capabilities at high acquisition speeds, absorbance measurements were performed on three distinct polymer samples: polystyrene (PS), polyethylene (PE), and polypropylene (PP). The absorbance spectra were baseline-corrected to account for Fresnel reflection and scattering losses by subtracting a low-order polynomial fit to the non-absorbing spectral regions. Figure~\ref{fig:Spectra_results} compares the resulting absorbance spectra to reference spectra obtained using a commercial ATR-FTIR spectrometer (Bruker, 4 cm$^{-1}$ resolution), each showing very close correspondence.

Figures~\ref{fig:Spectra_results}(a) and (b) show a comparison of a standard Perkin-Elmer polystyrene (PS) film spectra acquired at two different integration times, 10 ms (100 Hz) and 150 ms (6.6 Hz), respectively, and clearly illustrates the trade-off between acquisition speed and signal quality. At an integration time of 10~ms (100\,Hz) shown in Figure~\ref{fig:Spectra_results}(a), the system clearly resolves the aromatic C--H stretching modes at 3025 cm$^{-1}$ and the aliphatic C--H stretches between 2850 and 2924 cm$^{-1}$~\cite{liangInfraredSpectraHigh1958}. While the $\pm 1\sigma$ noise floor (shaded orange region) is visible at this speed, the spectral features remain distinct. Increasing the integration time to 150~ms, as seen inFigure~\ref{fig:Spectra_results}(b), significantly reduces the noise contribution, yielding a spectral profile where the central peak shape at 2924 cm$^{-1}$ becomes better resolved. A deformation in the peak profile is observed in this region in terms of both peak shape and its intensity, which can be attributed to signal saturation; here, the idler beam is strongly absorbed by the sample, and when combined with the intrinsic crystal absorption in this band (2850-2925 cm${^-1}$), the effective visibility approaches the noise floor. Nonetheless, the peak position remains distinguishable, allowing for excellent qualitative identification of the polymer in agreement with the reference ATR data.

Figure~\ref{fig:Spectra_results}(c) presents the spectrum of a 4~\textmu m thick polypropylene (PP) film acquired at 10~ms. Despite the short integration time, the three characteristic peaks associated with CH$_3$ and CH$_2$ vibrations (2953, 2918, and 2838 cm$^{-1}$) are clearly visible~\cite{andreassenInfraredRamanSpectroscopy1999}, confirming the system's ability to rapidly identify thin films. Finally, Figure~\ref{fig:Spectra_results}(d) shows the spectrum of a commercial polyethylene (PE) sample from a shopping bag, recorded at 10~ms. The characteristic C--H stretching doublet, comprising the asymmetric stretching mode at $\sim$2915 cm$^{-1}$ and the symmetric mode at $\sim$2850 cm$^{-1}$, is clearly distinguishable from the background noise~\cite{gulminePolyethyleneCharacterizationFTIR2002}. The ability to identify such commonplace materials shows the system's capacity for rapid plastic identification.

The successful retrieval of these polymer spectra confirms the suitability of the nonlinear interferometer module for mid-IR spectroscopy using room-temperature silicon detectors, achieving a single-shot SNR of 34 with a 10\,ms integration time. However, as observed in Figure~\ref{fig:Spectra_results}, the system's performance is currently constrained by its finite dynamic range, which leads to minor deviations in peak amplitude and shape in regions of strongest sample absorption.
This saturation effect stems from the baseline interference visibility, which averaged below 20\%  across the spectrum. The primary sources of this baseline reduction are the intrinsic absorption of the 30~mm long ppKTP crystal near 3.45~\textmu m (discussed in Section 3.1), combined with scattering and imperfect spatial-mode overlap. Consequently, in spectral regions where strong sample absorption of the idler light aligns with the intrinsic crystal absorption, the cumulative loss drives the signal amplitude toward the noise floor. This low SNR results in signal saturation and the observed peak distortions. Furthermore, while the interferometer was generally stable, residual pump intensity fluctuations introduced a minor noise contribution at longer time scales.
Nevertheless, despite these localized saturation effects, the system retains sufficient sensitivity to robustly identify the characteristic polymer signatures, highlighting its potential for real-world microplastics monitoring.

Addressing these limitations offers clear pathways for optimization. Utilizing shorter crystals or materials with reduced mid-infrared absorption would directly increase visibility and broaden the measurement's dynamic range. Furthermore, optimized focusing and mode matching, supported by high-quality anti-reflection coatings, would minimize losses. Beyond component optimization, a significant performance leap can be achieved by implementing a pump-enhanced cavity around the nonlinear crystal \cite{lindnerHighsensitivityQuantumSensing2023}. Resonant enhancement of the pump field could increase the photon-pair generation rate by up to 2 orders of magnitude. This would lead to a proportional boost in measurement speed at a fixed SNR, as extrapolated by the purple dashed line in Figure~\ref{fig:SNR_scaling_plot}. Such an enhancement would potentially enable acquisition rates up to 100 kHz, opening the door to fast scanning-based hyperspectral imaging. To support these advances while improving footprint and stability, future module iterations will include micro-integrated fibre coupling for the signal light, removing the need for additional external opto-mechanics. Furthermore, the existing fibre-coupled pump port already allows the laboratory-grade Ti:sapphire laser used here to be replaced with compact, low-cost diode lasers, paving the way for affordable and fully portable system implementations

\section{Conclusion}

In summary, we have demonstrated a modular nonlinear interferometer capable of retrieving mid-infrared absorption spectra using undetected photons. The system integrates the nonlinear crystal, Michelson-like geometry, and fibre-coupled pump into a compact and thermally stabilised module. This architecture ensures high mechanical stability, eases alignment, and provides direct compatibility with conventional silicon-based grating spectrometers. A total photon flux of $3.6 \times 10^{7}$ s$^{-1}$ was detected, yielding a single-shot SNR of 34 at 10 ms integration time. These performance metrics, combined with a spectral resolution of $\sim$6 cm$^{-1}$, confirm that the system operates near the shot-noise limit while maintaining sufficient resolution to distinguish  polymer absorption features.

The module successfully retrieved the characteristic vibrational bands of polypropylene, polyethylene, and polystyrene, validating its capability for broadband mid-infrared spectroscopy without the need for mid-IR detectors. While intrinsic absorption in the ppKTP crystal currently limits interference visibility, and consequently the dynamic range in specific bands, this is not a fundamental constraint of the undetected photon technique. Future iterations, with optimized crystal materials or alternative nonlinear media, could significantly extend the measurable spectral range and enhance interference visibility.

Looking ahead, the module offers a versatile platform for real-world sensing. Unlike conventional FTIR systems, our module contains no moving parts, inherently improving robustness for field applications. Such a system could be deployed for environmental monitoring, e.g in CO$_{2}$ emissions monitoring, microplastics analysis in water samples, or recycling quality control, or adapted with scanning optics for hyperspectral mid-IR microscopy in medical diagnostics. By combining compact module design, inexpensive detectors, and robust fibre coupling, undetected-photon spectroscopy provides a clear path toward accessible, scalable, and field-deployable detection technologies.

\begin{acknowledgments}
We acknowledge funding by the Bundesministerium für Forschung, Technologie und Raumfahrt via the grant Sim-QPla (FKZ 13N15943, 13N15944 ) and QEED (FKZ 13N16381, 13N16384) and EP further acknowledges  EPSRC-funded QuSIT (EP/Z633166/1). We thank Ivan Zorin for providing samples and ATR-FTIR comparison data.
\end{acknowledgments}

\appendix

\section{Appendix}

\subsection{Optimal crystal length}

\begin{figure}[htbp]
    \centering
    \includegraphics[width=\columnwidth]{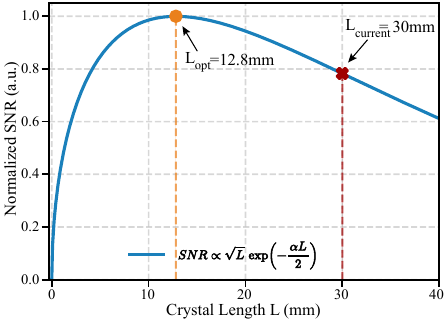}
    \caption{The figure shows the crystal length optimisation based on maximising the normalized SNR of the crystal with reference to its length and the inherit absorption of the KTP crystal. the red cross depicts the current value of this work. The orange circle shows the maximum achievable SNR with KTP as an SPDC source for the nonlinear interferometer.}
    \label{fig:crystal_optimzed}
\end{figure}

In the design of nonlinear interferometers, the crystal length is a critical parameter that dictates the trade-off between source brightness and interference visibility. To determine the optimal crystal length ($L_{\text{opt}}$) that maximizes the SNR, we model the system under shot-noise limited conditions assuming optimal focusing geometry. The scaling behavior is derived as follows:

\begin{enumerate}
\item \textbf{Brightness Scaling:} Under optimal focusing conditions (Boyd-Kleinman regime \cite{boydParametricInteractionFocused1968}), the generated photon flux ($\Phi$) scales linearly with the crystal length, $\Phi \propto L$. Consequently, the shot noise ($\sigma_{\text{shot}}$) associated with this flux scales as $\sqrt{\Phi} \propto \sqrt{L}$.
\item \textbf{Visibility Scaling:} The interference visibility ($V$) is attenuated by intrinsic absorption in the idler path. For an absorption coefficient $\alpha$, the transmission is $T = e^{-\alpha L}$. The visibility scales with the field amplitude ratio, proportional to $\sqrt{T} = e^{-\alpha L / 2}$.
\item \textbf{SNR Function:} The effective SNR is proportional to the product of the shot-noise-limited amplitude ($\Phi / \sigma_{\text{shot}} \propto \sqrt{L}$) and the visibility ($V$). Combining these yields the objective function:
\begin{equation}
    \text{SNR}(L) \propto \sqrt{L} \cdot e^{-\frac{\alpha L}{2}}
\end{equation}
\item \textbf{Optimization:} Differentiating with respect to $L$ and solving for the extremum:
\begin{align*}
    \frac{d(\text{SNR})}{dL} &\propto \frac{1}{2\sqrt{L}} e^{-\frac{\alpha L}{2}} - \frac{\alpha}{2}\sqrt{L} e^{-\frac{\alpha L}{2}} = 0 \\
    \frac{1}{2\sqrt{L}} &= \frac{\alpha \sqrt{L}}{2} \quad \Rightarrow \quad 1 = \alpha L \quad \Rightarrow \quad L_{\text{opt}} = \frac{1}{\alpha}
\end{align*}
\end{enumerate}

For the specific case of KTP at the idler wavelength of 3.45~\textmu m, the absorption coefficient is approximately $\alpha = 0.78 \text{ cm}^{-1}$ \cite{hanssonTransmissionMeasurementsKTP2000}. Substituting this value yields an optimal interaction length of:
\begin{equation*}
L_{\text{opt}} = \frac{1}{0.78 \text{ cm}^{-1}} \approx 1.28 \text{ cm} = 12.8 \text{ mm}.
\end{equation*}
This analysis indicates that the current 30~mm crystal is significantly longer than the optimal length for this specific absorption band. Future iterations should target a crystal length closer to $\sim$13~mm to maximise the SNR.


\bibliography{plastics}

\end{document}